**Evolution of Magnetism in Single-Crystal $Ca_2Ru_{1-x}Ir_xO_4$ ($0 \le x \le 0.65$)**

S. J. Yuan[1*], J. Terzic[1], J. C. Wang[1,2,3], S. Aswartham[1], W. H. Song[1,4], F. Ye[2], and G. Cao[1*]

[1]*Center for Advanced Materials, Department of Physics and Astronomy, University of Kentucky, Lexington, Kentucky 40506, USA*

[2]*Quantum Condensed Matter Division, Oak Ridge National Laboratory, Oak Ridge, Tennessee 37831, USA*

[3]*Department of Physics, Renmin University of China, Beijing 100872, China*

[4]*Institute of Solid State Physics, Chinese Academy of Sciences, Hefei 230031, China*

# ABSTRACT

We report structural, magnetic, transport and thermal properties of single-crystal $Ca_2Ru_{1-x}Ir_xO_4$ ($0 \le x \le 0.65$). $Ca_2RuO_4$ is a structurally-driven Mott insulator with a metal-insulator transition at $T_{MI}$ = 357 K, which is well separated from antiferromagnetic order at $T_N$ = 110 K. Substitution of 5d element, Ir, for Ru enhances spin-orbit coupling (SOC) and locking between the structural distortions and magnetic moment canting. In particular, Ir doping intensifies the distortion or rotation of $Ru/IrO_6$ octahedra and induces weak ferromagnetic behavior along the ***c***-axis. Moreover, the magnetic ordering temperature $T_N$ increases from 110 K at $x = 0$ to 215 K with enhanced magnetic anisotropy at x = 0.65. The effect of Ir doping sharply contrasts with that of 3d-element doping such as Cr, Mn and Fe, which suppresses $T_N$ and induces unusual negative volume thermal expansion. The stark difference between 3d- and 5d-element doping underlines a strong magnetoelastic coupling inherent in the Ir-rich oxides.

PACS numbers: 71.70.Ej, 75.30.Kz, 71.30.+h

## I. INTRODUCTION

The Coulomb interaction *U* is generally comparable to the 4d-bandwidth *W* in the 4d-based ruthenates, which leaves them precariously balanced on the border between metallic and insulating behavior, or on the verge of long-range magnetic order. A common characteristic of these materials is that underlying physical properties are critically linked to the lattice and orbital degrees of freedom and tend to exhibit a giant response to modest lattice changes. This is dramatically illustrated by $Sr_2RuO_4$ and $Ca_2RuO_4$, where the former compound exhibits a prototypical *p*-wave superconducting state [1] that strongly contrasts with the more distorted structure (due to a smaller ionic radius $r_{Ca} < r_{Sr}$ ) and first-order metal-insulator (MI) transition observed for the latter compound [2, 3].

Extensive investigations of $Ca_2RuO_4$ [4, 5] have established that a strong cooperative Jahn-Teller distortion removes the degeneracy of the three Ru $t_{2g}$ orbitals ($d_{xy}$, $d_{yz}$, $d_{zx}$) via a transition to orbital order that, in turn, drives the MI transition at $T_{MI}$ = 357 K [6-14]. Classic Mott insulators undergo simultaneous transitions to antiferromagnetic (AFM) order and an insulating state at $T_{MI}$. However, $Ca_2RuO_4$ undergoes AFM order at $T_N$ = 110 K ≪ $T_{MI}$,[2] and is therefore a highly interesting and unique archetype of a MI transition that is strongly coupled to a structural transition and is not driven by AFM exchange interactions.

We recently observed that slight substitutions of a 3d element M (M = Cr, Mn, Fe) for Ru shifts $T_{MI}$, weakens the orthorhombic distortion, and induces either metamagnetism or magnetization reversal below $T_N$.[12-14] Furthermore, M doping for Ru produces substantial negative thermal expansion in $Ca_2Ru_{1-x}M_xO_4$, with a total volume expansion ratio $\Delta V/V$ as high as 1% on cooling. The onset of the negative thermal expansion closely tracks shifts of $T_{MI}$ and $T_N$, and sharply contrasts with classic examples of negative thermal expansion that show no correlation with electronic properties. These unusual observations suggest a complex interplay between orbital, spin, and lattice degrees of freedom.[12-14]

It is important to note $Ru^{4+}(4d^{4+})$ ions tend to adopt a low-spin state or S = 1

state because relatively large crystal fields often overpower the Hund's rule coupling.[15] On the other hand, the spin-orbit coupling (SOC) may be strong enough to impose a competing singlet, or an angular momentum $J_{\mathrm{eff}} = 0$, ground state.[15, 16] Compared to 4d-ruthnates, 5d-iridates have stronger SOC (~0.4 eV, compared to ~0.16 eV for Ru ions)[17], which compete vigorously with Coulomb interactions, non-cubic crystalline electric fields, and Hund's rule coupling. [17-21] A profound manifestation of this competition is the novel "$J_{\mathrm{eff}} = 1/2$ Mott state" that was recently observed in the layered iridates with tetravalent $Ir^{4+}(5d^{4+})$ ions.[18, 19] Therefore, substitutions of Ir for Ru in 4d-ruthnates is expected to promote novel magnetic behavior. Moreover, in light of the novel insulating state recently discovered in $Sr_2IrO_4$,[18], a comparison with its isostructural compound $Ca_2IrO_4$ would be desirable. However, the structural instability prevents the formation of the perovskite-like $Ca_2IrO_4$; the heavily Ir-doped $Ca_2RuO_4$ or $Ca_2Ru_{1-x}Ir_xO_4$ with x up to 0.65 thus provides an alternative for comparison and contrast to the archetype $J_{\mathrm{eff}} = 1/2$ insulator $Sr_2IrO_4$ that antiferromagnetically orders at $T_N$=240 K.[22]

In this paper, we report results of our study of single-crystal $Ca_2Ru_{1-x}Ir_xO_4$ with $0 \le x \le 0.65$. Our central findings are that increasing Ir substitution induces a dramatic increase in moment canting and the appearance of a weak ferromagnetic (FM) moment along the ***c***-axis. The magnetic ordering temperature $T_N$ increases from 110 K at $x = 0$ to 215 K at x = 0.65, along with enhanced magnetic anisotropy due to increased SOC. The increase in both $T_N$ and $T_{MI}$ with increased Ir doping closely follows the enhanced Ru/$IrO_6$ octahedral rotation or reduced Ru/Ir-O-Ru/Ir bond angle. This study reveals that Ir doping enhances the coupling between the lattice and magnetic moment, sharply contrasting 3d element doping that readily reduces such a coupling and orthorhombic distortions, thus suppresses the AFM and insulating states. The pronounced difference illustrated in this study highlights a strong magnetoelastic coupling inherent in the SOC-driven iridates that dictates magnetic properties. This work also provides an important comparison to the extensively studied $Sr_2IrO_4$.

## II. EXPERIMENTAL

Single crystals were grown using flux techniques described elsewhere.[23] The structures of $Ca_2Ru_{1-x}Ir_xO_4$ were determined using a Nonius Kappa CCD x-ray diffractometer at 90 K. Structures were refined by full-matrix least squares using the SHELX-97 programs.[24] All structures affected by absorption and extinction were corrected by comparison of symmetry-equivalent reflections using the program SADABS.[24] It needs to be emphasized that the single crystals are of high quality and there is no indication of any mixed phases or inhomogeneity in the single crystals studied. The standard deviations of all lattice parameters and interatomic distances are smaller than 0.1%. Chemical compositions of the single crystals were estimated using both single-crystal x-ray diffraction and energy dispersive X-ray analysis (Hitachi/Oxford 3000). Magnetization, specific heat and electrical resistivity were measured using either a Quantum Design MPMS-7 SQUID Magnetometer and/or a Physical Property Measurement System with 14-T field capability.

## III. RESULTS AND DISCUSSION

$Ca_2RuO_4$ adopts a very peculiar distortion of the $K_2NiF_4$-prototype with a *Pbca* (61) space group consisting of layers of $RuO_6$ octahedra separated by Ca atoms.[4, 14] Neighboring corner-shared octahedra tilt and rotate in an ordered manner, as a result, the Ru-O-Ru bond angle is severely distorted from $180^o$.

Substituting $Ir^{4+}$ for $Ru^{4+}$ preserves the crystal structure but results in a reduction in the ***a***- and ***b***-axis lattice parameters and an elongation in the ***c***-axis lattice parameter, and eventualy shrinks the unit cell volume *V*, as shown in **Fig.1 (a)**. Compared to the parent compound $Ca_2RuO_4$, the $c/a$ ratio increases, by1.9% for $x$ = 0.5 at 90 K, for example. The orthorhombic distortion also increases with increasing x; e.g., $(b-a)/b$ = 0.0437465 for x = 0, and $(b-a)/b$= 0.0467 for x = 0.5 at 90 K. The decrease in the bond angle Ru/Ir-O1-Ru/Ir, Θ, is a further manifestation of more distorted $Ru/IrO_6$ octahedra for the Ir-doped compounds, sharply contrasting with that for a 3d element doping.[12-14] The decrease in Θ has important implications for magnetic and transport properties, as discussed below.

The magnetic susceptibility $\chi$(T) of the parent compound $Ca_2RuO_4$ exhibits a sharp anomaly due to AFM ordering at $T_N$=110 K (see inset in **Fig.2a**).[2] Ir doping induces three pronounced changes in the magnetic properties of single-crystal $CaRu_{1-x}Ir_xO_4$, as shown in **Fig. 2**. First, Ir substitution induces a weak FM behavior along ***c***-axis, and a notable magnetic anisotropy evident in **Fig. 2a**. The sizable hysteresis in isothermal magnetization along the ***c***-axis is consistent with the weak FM behavior, as shown in **Fig. 2b**. Second, Ir doping significantly increases the magnetic ordering temperature $T_N$ along the ***c***-axis, from 110 K at $x$ = 0 to 190 K at $x$ = 0.34, as shown in **Fig. 2a**. Third, there is another anomaly below $T_N$, denoted by $T_{SR}$ in Fig. **2a**, which is related to the spin-reorientation of the Ru/Ir ions.

$Ca_2RuO_4$ has a canted AFM structure adapted to a Dzyaloshinskii-Moriya (D-M) interaction on a distorted orthorhombic perovskite structure.[14, 25-28] The spins are canted away from the ***ab***-plane toward the ***c***-axis; consequently the value of the susceptibility along the ***ab***-plane is lower than along the ***c***-axis (see inset in **Fig. 2a**). The crystal and magnetic structures suggest that the easy axis for AFM order lies in the ***ab***-plane.[2, 4] The susceptibility cusp at $T_N$=110 K indicates that the canted moments in successive layers interact antiferromagnetically. The enhanced distortions in Ir-doped compounds $CaRu_{1-x}Ir_xO_4$ having larger *c*/*a* ratios and smaller Ru/Ir-O1-Ru/Ir bond angles further reduces the symmetry and enhances the D-M interaction. In contrast to the parent compound $Ca_2RuO_4$, the interlayer interaction in Ir-doped compounds drives the weak FM behavior observed along the ***c***-axis (see **Figs**. **2** and **3**). **Fig. 2c** shows a schematic picture of the moment configuration of $CaRu_{1-x}Ir_xO_4$. The net moments along the ***c***-axis for individual layers exhibit FM coupling due to canting. It is remarkable that the interlayer coupling changes from AFM coupling for the parent compound $Ca_2RuO_4$ to FM coupling for Ir-doped compounds.

Indeed, the evolution of the magnetic behavior is remarkably consistent with a theoretical proposal for the iridates that suggests an increased *c/a* ratio tends to result in a spin-flop transition to a collinear magnetic order along the ***c***-axis due to a strong magnetoelastic coupling.[29] That the increase in $T_N$ closely tracks the decrease in the Ir/Ru-O1-Ir/Ru bond angle Θ also manifests the strong magnetoelastic coupling (**Fig.1b**).

It is now recognized that the 5d-based iridates have strong SOC that competes vigorously with Coulomb interactions, non-cubic crystalline electric fields, and other relevant energies, leading to the "$J_{eff}$ = 1/2 state".[17-21] One profound result of this competition is that 5d-iridates exhibit complex magnetic states with high critical temperatures, such as $Sr_2IrO_4$ ($T_N$ = 240 K)[22], $Sr_3Ir_2O_7$ ($T_N$ = 285 K)[30, 31] and $BaIrO_3$ ($T_C$ = 183 K)[32, 33]. It is established that the magnetic moment and ordering temperature are closely associated with the Ir-O-Ir bond angle Θ.[34] In particular, a recent study reveals that there are a perfect locking between the octahedral rotation and magnetic moment canting angles that can persist even in the presence of large noncubic local distortions.[34, 35] Since Ir doping further reduces Θ, it is not surprising that $T_N$ steadily rises with $x$, as shown in **Figs. 3a and b**; $T_N$ reaches 215 K for 65% of Ir doping, and would approach 270 K for 100% of Ir doping or $Ca_2IrO_4$ according to the upward trajectory in **Fig.3c** should perovskite-like $Ca_2IrO_4$ exist.

The temperature dependence of the electrical resistivity $\rho(T)$ of $CaRu_{1-x}Ir_xO_4$ is shown in **Figs. 4a**. It is clear that the MI transition increases from $T_{MI}$ = 357 K for $x$ = 0 to $T_{MI}$ = 369 K for $x$ = 0.016, and $T_{MI}$ = 384 K for $x$ = 0.03, beyond which it is no longer well defined. The increase in $T_{MI}$ closely tracks the enhanced distortions of the Ru/$IrO_6$ octahedra with reduced Ru/Ir-O1-Ru/Ir bond angle Θ. This behavior contrasts with that of a 3d transition-metal ion M (Cr, Mn, Fe) that weakens the orthorhombic distortions, thus insulating state.[12, 13]. The resistivity data over the interval 220 < $T$ < 290 K fit an activated behavior with a gap of about 0.40 eV for x = 0, and 0.28 eV for the Ir doped crystals. Variable-range hopping (VRH) model ($\rho \sim \exp(1/T)^{1/2}$) fits were more successful for x = 0, suggesting Anderson localization is relevant in the parent compound. However, VRH fails to describe the resistivity of Ir-doped crystals.

The heat capacity $C(T)$ data for $0.016 \leq x \leq 0.65$ show weak or no anomaly at $T_N$, while the anomaly around $T_{MI}$ for x = 0.016 confirms the MI transition.[13] Fitting the data to $C(T) = \gamma T + \beta T^3$ for 1.7 < $T$ < 20 K yields the Sommerfeld coefficient $\gamma$ for the electronic contribution to $C(T)$, which serves as a measure of the electronic density of states at the Fermi level, $N(E_F)$ and the effective mass of the carriers. There is no substantial increase of $\gamma$ with Ir concentration, as shown in **Figs. 4b**. The small values of $\gamma$ are consistent with the low electrical conductivity observed at low temperatures. The slight increase in $\gamma$ with increasing $x$ results from the moderate drop in activation gap for Ir-doped compounds.

## IV. CONCLUSIONS

The substitution of Ir for Ru in $CaRuO_4$ enhances the SOC and intensifies the distortions of the Ru/$IrO_6$ octahedra. As a result, the MI transition rises and a pronounced weak ferromagnetic behavior occurs, which strengthens with increasing Ir concentration. The magnetic ordering temperature $T_N$ increases from 110 K at $x = 0$ to 215 K at x = 0.65, which is remarkably comparable to 240 K for $Sr_2IrO_4$,[22] along with enhanced magnetic anisotropy due to SOC. The increase in both $T_N$ and $T_{MI}$ with increased Ir doping closely follows the enhanced Ru/$IrO_6$ octahedral rotation or reduced Ru/Ir-O1-Ru/Ir bond angle. More generally, the effect of Ir doping tends to strengthen the coupling between the lattice and magnetic moment whereas a 3d element doping readily reduces such a coupling and the orthorhombic distortions, thus suppresses the AFM and insulating states and causes the unusual negative volume expansion as well. For comparison and contrast, the magnetic susceptibility for some representative 3d-element and Ir-doped $Ca_2RuO_4$ samples is illustrated in **Fig.5**, where $T_N$ drastically increases up on Ir doping but decreases due to Cr or Fe doping. The sharp contrast highlights a strong magnetoelastic coupling or locking between the octahedral rotation and magnetic moment canting angles, a pronouced characteristic of the SOC-driven iridates such as $Sr_2IrO_4$, $Sr_3Ir_2O_7$ and $BaIrO_3$.[17, 22, 30, 32, 34, 35]

## ACKNOWLEDGMENTS

This work was supported by the National Science Foundation via Grant No. DMR-1265162.

*Corresponding authors: shujuan.yuan@uky.edu; cao@uky.edu

## REFERRENCES

**Figure Captions**

FIG. 1. Ir concentration $x$-dependence of (a) the ***a***-, ***b***-, and ***c***-axis lattice parameters (right scale); and (b) Ru/Ir-O1- Ru/Ir bond angle Θ for $CaRu_{1-x}Ir_xO_4$ at $T$ = 90 K. Inset: Schematics of the distorted Ru/Ir-O1-Ru/Ir bond angle Θ.

FIG. 2. (a) Magnetic susceptibility $\chi(T)$ at $\mu_0 H = 0.1$ T, (b) the isothermal magnetization M(H) at 1.7 K, for x = 0.34. The inset in (a) shows $\chi(T)$ for x = 0. The $\chi(T)$ data were measured under 0.1 T after field cooling (FC). The *M(H)* were measured after a zero-field-cooled (ZFC) process. (c) The schematic magnetic-structure derived from the magnetic results for $CaRu_{1-x}Ir_xO_4$.

FIG. 3. Representative magnetic susceptibilities $\chi(T)$ in the ***ab***-plane (a) and along the ***c***-axis (b) after field-cooling in an applied field $\mu_0 H = 0.1\,T$ for $CaRu_{1-x}Ir_xO_4$ with $x$ = 0.28, 0.34, 0.50, and 0.65; the Ir concentration $x$ dependence of Neel temperature $T_N$ (c). The data are derived from $\chi(T)$ data for field along the ***c***-axis.

FIG. 4. Temperature dependence of the resistivity $\rho(T)$ in the ***ab***-plane (a) for representative compositions x = 0, 0.016, 0.03, and 0.5. The inset in (a) illustrates variable range hoping (VRH) in a plot of $\ln\rho_a$ vs $T^{-1/2}$ for x = 0 and 0.50. (b) Sommerfeld coefficient $\gamma$ vs $x$, for $CaRu_{1-x}Ir_xO_4$.

FIG. 5. Magnetic susceptibility as a function of temperature for some representative 3d-element and Ir-doped $Ca_2Ru_{1-x}M_xO_4$ compounds including the parent compound $Ca_2RuO_4$.

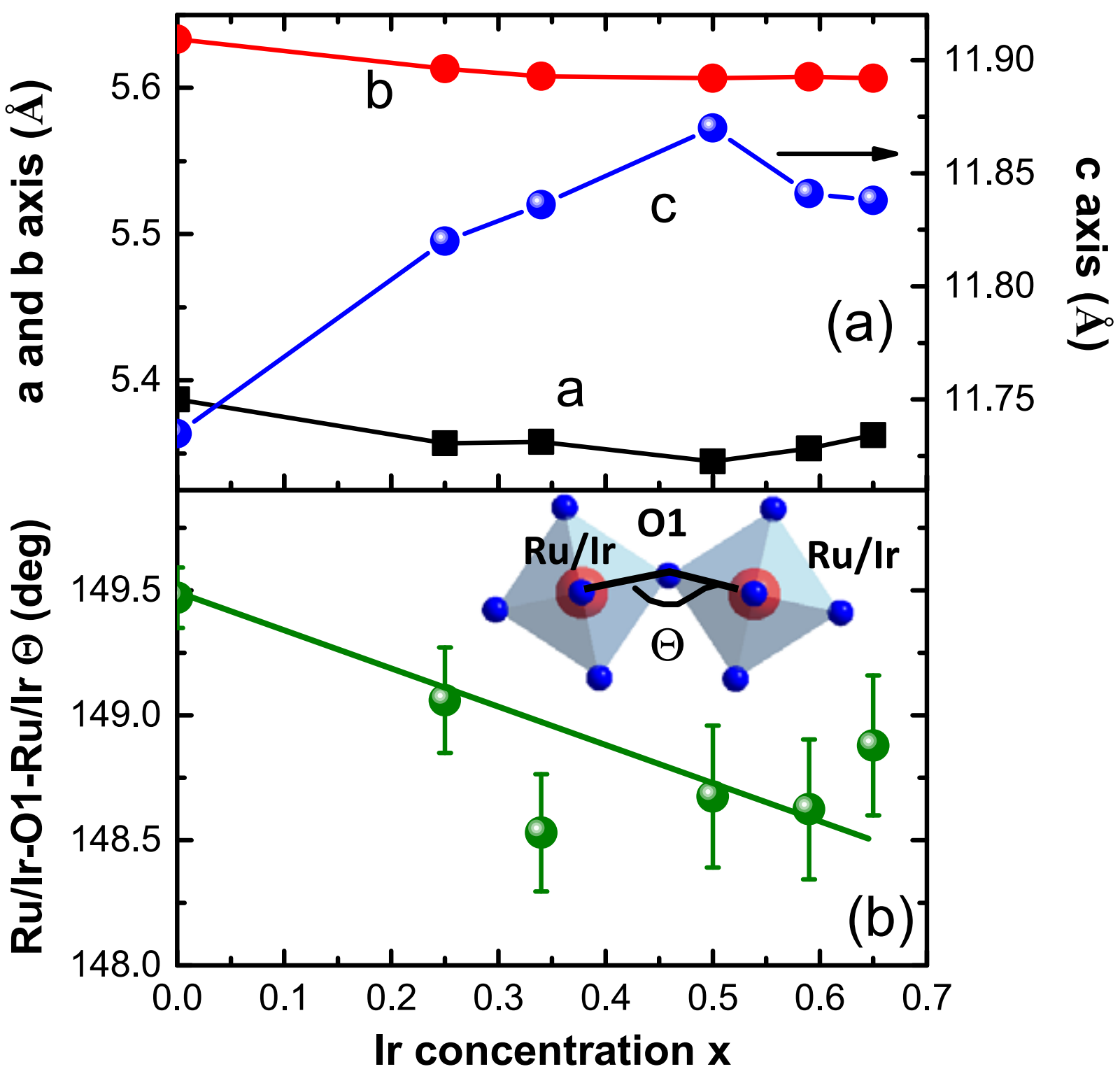

**Fig. 1**

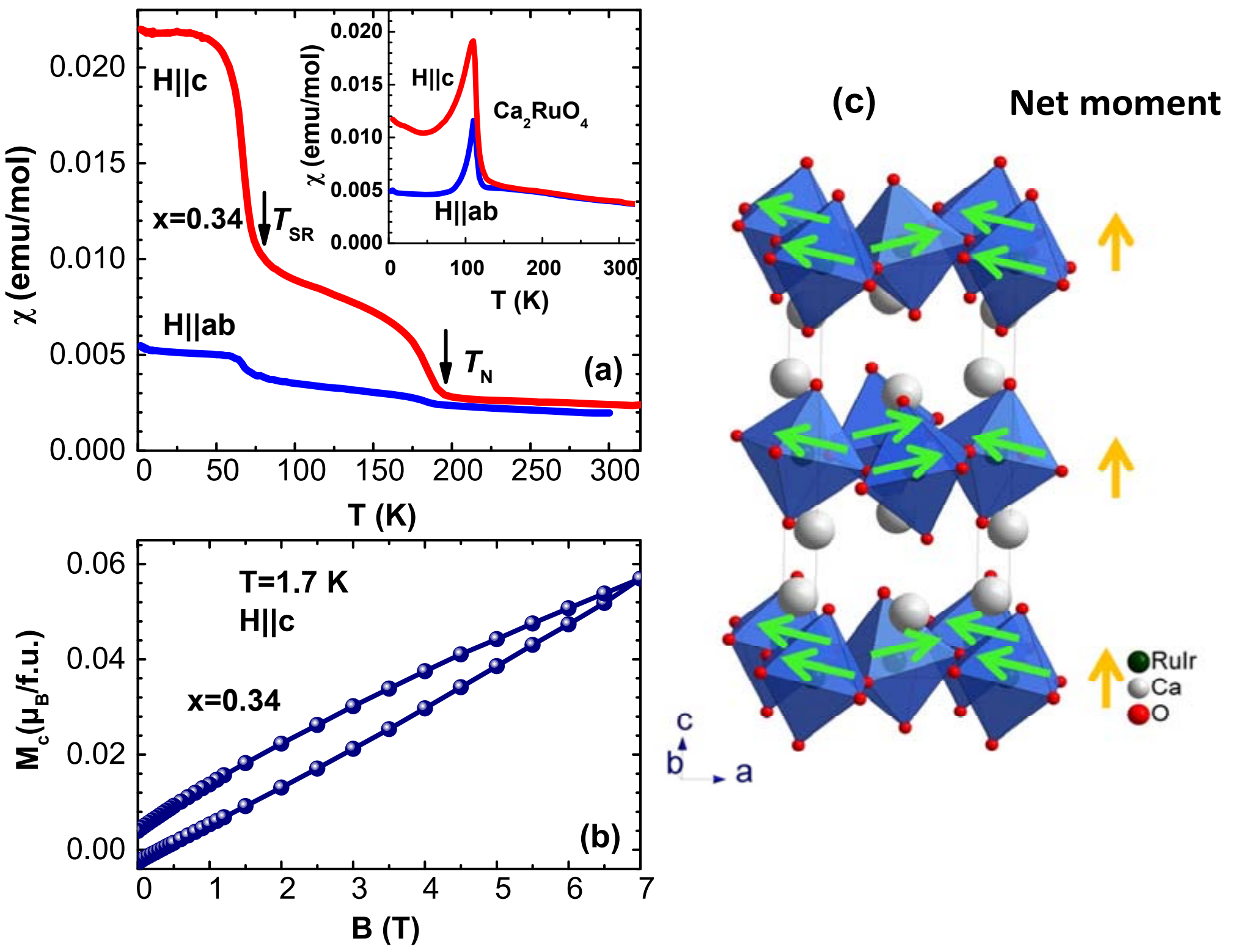

**Fig.2**

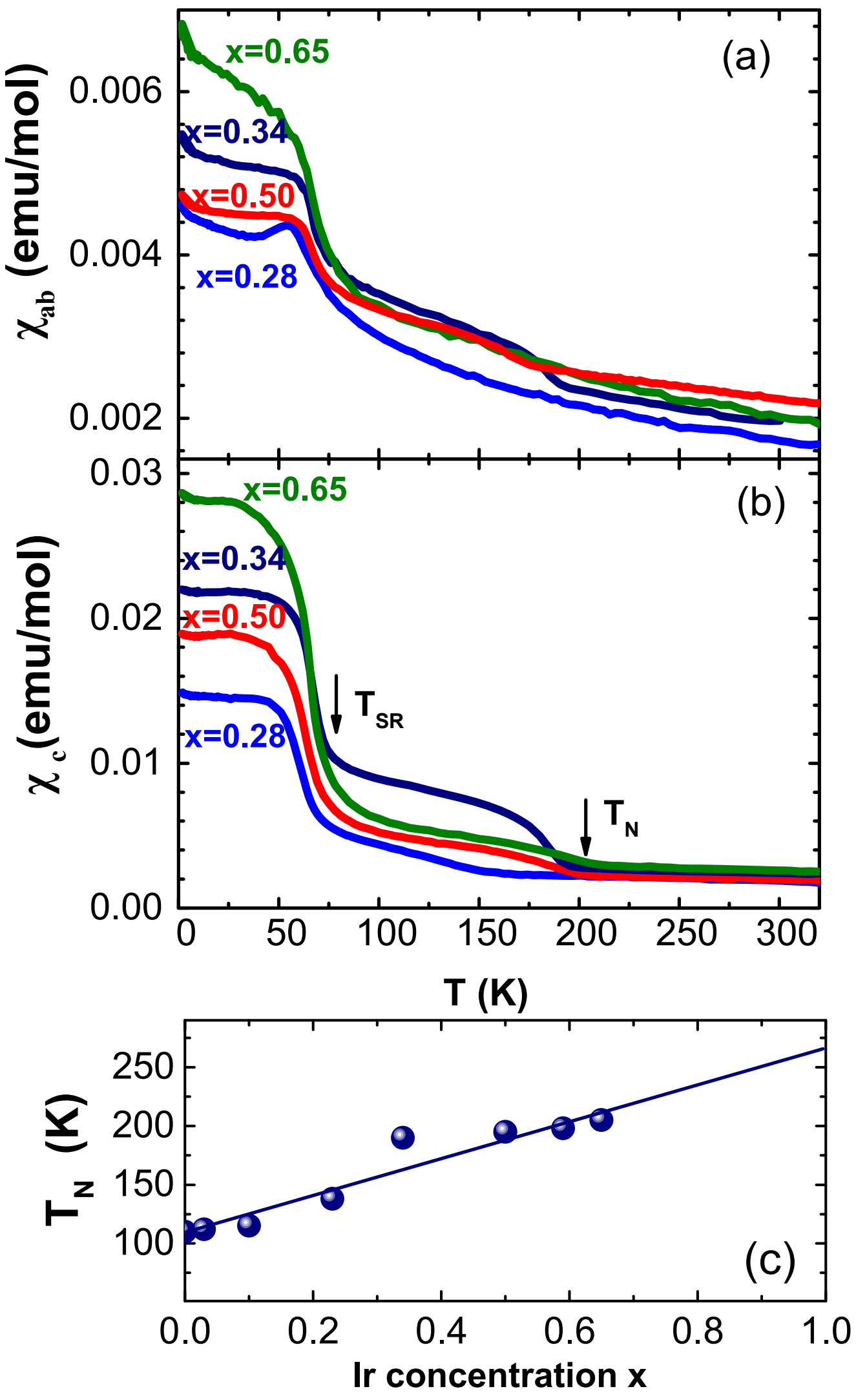

**Fig. 3**

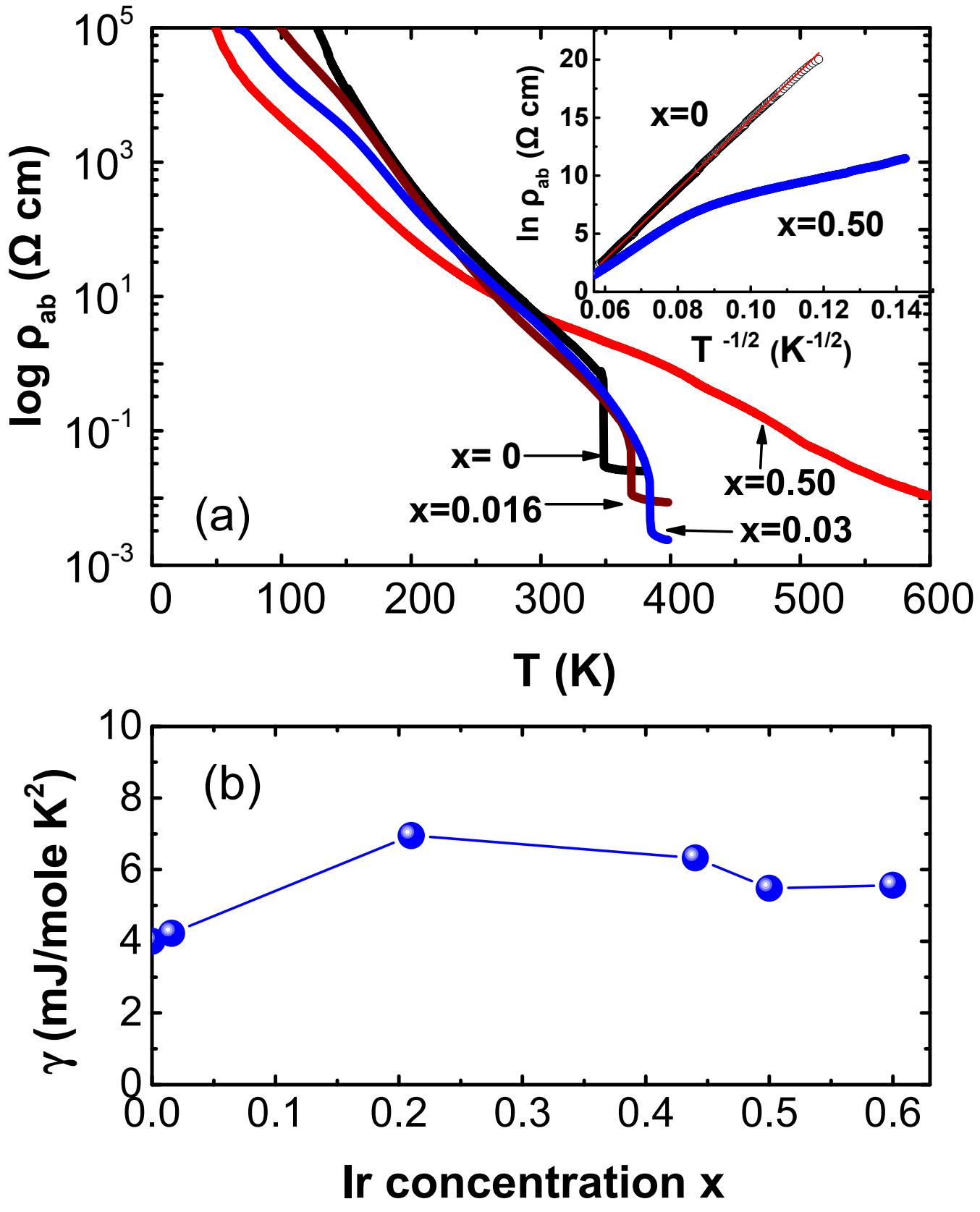

**Fig. 4**

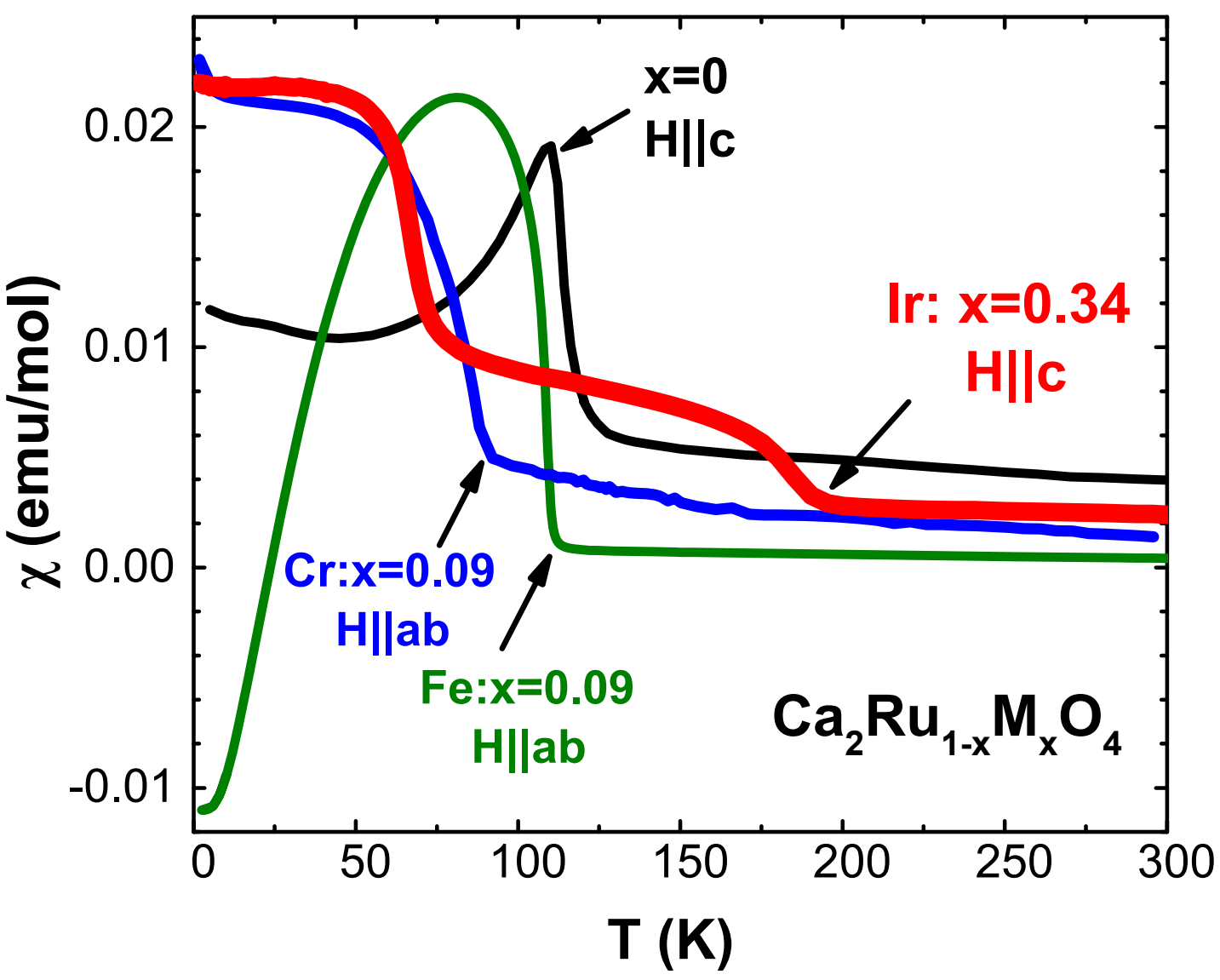

**Fig. 5**